\documentclass[preprint]{aastex}

\usepackage{amsmath,amssymb}

\usepackage{graphicx}

\usepackage{natbib}

\begin{document}

\title{SEP acceleration in CME driven shocks using a hybrid code}

\author{L. Gargat\text{\'e}\altaffilmark{1}\footnote{Present address\: Critical Software S.A. Parque Industrial de Taveiro, Lote 49, 3045-504 Coimbra, Portugal.},  R. A. Fonseca\altaffilmark{1,}\altaffilmark{2}, L. O. Silva\altaffilmark{1}.}
\affil{\altaffilmark{1}GoLP/Instituto de Plasmas e Fus$\tilde{a}$o Nuclear, Instituto Superior T\text{\'e}cnico, 1049-001 Lisboa Portugal.}
\affil{\altaffilmark{2}{ISCTE - Instituto Universit{\'a}rio de Lisboa, 1649-026 Lisboa, Portugal.}}
\author{R.A. Bamford\altaffilmark{3}}
\affil{\altaffilmark{3}Rutherford Appleton Laboratory, Chilton, Didcot, OX11 0QX, UK.}
\author{R.Bingham\altaffilmark{4,}\altaffilmark{3}}
\affil{\altaffilmark{3} SUPA, University of Strathclyde, Glasgow, Scotland, 4G 0NG,U.K.}

\email{Ruth.Bamford@stfc.ac.uk}

		\date{\hspace{2.5in}June 18th 2014}

\begin{abstract}

We perform hybrid simulations of a super-Alfv{\'e}nic quasi-parallel shock, driven by a Coronal Mass Ejection (CME), propagating in the Outer Coronal/Solar Wind  at distances of between 3 to 6 solar radii. The hybrid treatment of the problem enables the study of the shock propagation on the ion time scale, preserving ion kinetics and allowing for a self-consistent treatment of the shock propagation and particle acceleration. The CME plasma drags the embedded magnetic ﬁeld lines stretching from the sun, and propagates out into interplanetary space at a greater velocity than the in-situ solar wind, driving the shock, and producing very energetic particles. Our results show that electromagnetic Alfv{\'e}n waves are generated at the shock front. The waves propagate upstream of the shock and are produced by the counter-streaming ions of the solar wind plasma being reflected at the shock. A significant fraction of the particles are accelerated in two distinct phases\: first, particles drift from the shock and are accelerated in the upstream region and, second, particles arriving at the shock get trapped, and are accelerated at the shock front. A fraction of the particles diffused back to the shock, which is consistent with the Fermi acceleration mechanism. 

\end{abstract}

\keywords{Acceleration of particles, Sun: coronal mass ejections (CMEs), Sun: particle emission, Shock waves, (Sun:) solar–-terrestrial relations, plasmas} 


\maketitle

\section{INTRODUCTION}

Coronal Mass Ejections (CME) are large ejections of solar material that periodically erupt from the Sun \cite[]{Gopalswamy2003, forbes2006cme, Gopalswamy2008}. As CMEs propagate out into interplanetary space, they can produce transient bursts of extremely energetic particles referred to as Solar Energetic Particle (SEP) events \cite[]{Sheeley1983, Kahler2001, Gopalswamy2003, Kahler2004}. 

To be identified as an SEP the flux of particles (protons, electrons with trace higher Z ions) with energies above 10 MeV, must be greater than 10pfu (particle flux units $=$ particles per cm$^{-2}$sec$^{-1}$ str$^{-1}$) \cite[]{Gopalswamy2003}. 

The energy spectra of the SEP populations varies considerably \cite[]{Lin1974, Hollebeke1975, Kallenrode1992} and shows a dependence on the associated parameters of the originating CME \cite[]{Park2012}. Often the observed particle energies reach several hundred MeV \cite[]{Reames1999}, and even GeV energies \cite[]{Ryan2000, Kim2013}. SEP events can last from a period of several hours, up to several days \cite[]{Reames1999}. The combination of high flux and high penetrating particles mean that SEP events intersecting the Earth and man-made technology in space, present a significant ``Space Weather" risk of damage and disruption to vulnerable systems \cite[]{Feynman2000} and human tissue of astronauts \cite[]{Wu2011}. The SEPs from CME shock events tend to be the more extended in duration, or `gradual events', and have the harder energetic particle spectrum \cite[]{Kahler2001, Kahler2004, Kahler2003} and so the most interest for Space Weather mitigation.

The characteristics of high flux and high energy spectra suggest a very effective acceleration mechanism associated with CME shock. While acknowledging that SEP-type events maybe associated with other phenomena \cite[]{Tylka2006, Rouillard2012}, here we consider the acceleration mechanism of CMEs propagating faster than $\sim 800\,km\,s^{-1}$ \cite[]{Gopalswamy2008}. At these propagation speeds, if the local plasma density $n$ and magnetic field strength $B$ encountered by the CME  are such that the wave front is travelling super-Alfv{\'e}nically, then it will create an interplanetary shock \cite[]{Gopalswamy2003, Park2012}. Correlations between CME parameters of linear speed, angular extent and relative location on the Sun, have shown the greatest predictor of SEP event occurrence and particle flux goes with increasing CME speed, 30\% for 800 kms$^{-1}$ to 100\% for CME speeds of 1800 kms$^{-1}$ \cite[]{Gopalswamy2008, Hwang2010,Park2012}. The presence of preceding CMEs has also been found to decide the peak solar energetic particle flux \cite[]{Gopalswamy2003, Gopalswamy2004} further indicating that the important parameter is the local Alfv{\'e}n speed which is being reduced ahead of the second CME front resulting in harder SEP spectra.

Current particle acceleration mechanisms from collisionless shocks \cite[]{sagdeev1966} include shock drift acceleration  and diffusive shock acceleration. The shock drift acceleration mechanisum, dominant for perpendicular shocks, was originally studied by \cite[]{Dorman1959, Schatzmann1963}; more recent reviews \cite[]{Decker1983, Toptygin1983} estimate that the maximum energy gains attainable are $\sim 5 \times$ the initial particle energies, and depend on the magnetic field compression ratio due to the shock. The diffusive shock acceleration mechanism (or first order Fermi acceleration) \cite[]{Bell1978, Blandford1978} is thought to be responsible for the highest-energy particles observed at quasi-parallel shocks, thus being the preferred mechanism for Cosmic Ray acceleration, and also being used to explain some features of particle spectra from SEP events. In diffusive shock acceleration, particles crossing the shock front are accelerated by successive reflections downstream and upstream due to turbulence, potentially reaching very high energies. Fundamental theory on shock acceleration can be found in \cite[]{Toptygin1983, Stone1985,Volk1987, Jones1991}. 

Although turbulence exists in the solar wind for particle reflection to occur, its level is not sufficient to explain the production of MeV and GeV particles during the time CMEs and Interplanetary shocks take to reach the Earth \cite[]{sagdeev1991}. Instead, turbulence is produced at the shock by waves arising at the shock front and propagating upstream \cite[]{McKenzie1982, Gordon1999, Zank2000, Ng2003, Rice2003}.

In this paper we use a kinetic ion/fluid electron numerical simulation approach commonly known as a hybrid code \cite[]{dawson1983pic, fonseca2002} to study the propagation of a quasi-parallel CME shock in the solar wind environment. The code was originally developed to study the interaction of artificial plasmas released in the solar wind \cite[]{bingham1991simulation, gargate2008} and is now a massively parallel 3D hybrid particle code, dHybrid, \cite[]{gargate2007}, to simulate the solar wind environment and the acceleration mechanisms of solar energetic particles. The code has been successfully used to investigate cosmic ray  acceleration at collisionless shocks \cite[]{gargate2012}. The hybrid model uses massless fluid electrons and kinetic ions. The parallel implementation of this model allows the study of large regions of space (e.g. hundreds of ion gyro radius) over extended periods of time (e.g. tens of ion gyro periods), ideal for space plasma studies. Here we consider a CME driving a fast magnetosonic shock, with shock parameters known to correlate well with SEP events \cite[]{Park2012}. In our simulations a CME structure propagates at speeds of up to 1000 km/s interacting with the slower solar wind. The interactions cause the formation of a large scale quasi-parallel shock structure due to the flowing CME. The acceleration mechanisms of high energy particles are studied in this scenario. In the early acceleration phase, our results show that particles crossing the shock front accelerate perpendicularly to the shock front while maintaining their parallel velocity, supporting a surfatron-like acceleration model. The importance of this acceleration model as a means of providing a seed particle population for further acceleration is studied.

We explore the scenario of SEP acceleration and wave formation at CME driven quasi-parallel shocks using a hybrid model; the shock evolution can be followed on the ion time scale, the ion acceleration at the shock front is correctly modelled, and the smaller electron time scales can be neglected by using an ideal fluid model for this species. 

In comparison with MHD simulations, which do not capture kinetic effects and follow the evolution of a CME on a global scale, and over a time period relevant for the propagation of a CME in interplanetary space, hybrid simulations are localized in space, modelling a small part of the CME shock front and running over a time period relevant for the ion dynamics. 

Results from dHybrid show the self-consistent formation of Alfv\`{e}n waves upstream of the shock, with turbulence building up due to wave breaking, and strong particle acceleration. Energy gains of up to 110 times the maximum possible energy gain in one shock crossing are measured.

For the most accelerated particles, the observed energy gain is approximately quadratic in time, during the simulation time frame, consistent with surfatron acceleration \cite[]{katsouleas1983, ucer2001unlimited, lee1996pickup}, while for another less energetic set of particles the energy scales with $t^{1/2}$ consistent with diffusive shock acceleration. The observed energy gain would allow for a typical solar wind proton to reach an energy of hundreds of MeV in some minutes. A thorough discussion about the observed acceleration mechanisms will be presented.

This paper is organized as follows. In the next section, we present the numerical model in detail, describe the simulation setup, and present the plasma parameters assumed. In the Results section, we investigate the wave formation, the wave-particle interaction mechanisms, and particle acceleration. We also include a simple single-particle theoretical model that clarifies how particles are accelerated in the upstream Alfv{\'e}n waves, consistent with the observed simulation results. Finally, in the last section, we present the conclusions. 

\section{NUMERICAL MODEL}
\subsection{The hybrid model}

Hybrid models, with kinetic ions and fluid electrons, are commonly used in many problems in plasma physics \cite[]{Lipatov2002}. While MHD simulations are used to model CMEs globally, we use hybrid simulations to study shock properties locally, providing a new perspective over the problem of particle acceleration in gradual events.

The hybrid set of equations is derived neglecting the displacement current in Amp{\'e}re’s Law, considering quasi-neutrality and calculating moments of the Vlasov equation for the electrons in order to obtain the generalised Ohm’s Law. In our implementation of the hybrid model in the massively parallel three-dimensional (3D) code dHybrid \cite[]{gargate2007}, the effects of electron mass, resistivity and electron pressure are not considered; thus, the electric field is simply given by $\vec{E} = -\vec{V_e} \times \vec{B}$ ,  which can also be expressed as
\begin{equation}
          \vec{E} = -\vec{V_i} \times \vec{B} + \frac{1}{n} \left ( \nabla \times \vec{B} \right ) \times \vec{B}\label{Equ1}
\end{equation}

where we have used $\vec{V_e} = - \vec{J}/(|e|n) \times \vec{V_i}$, where $\vec{V_i} = \frac{1}{n} \int f_i \vec{v}\, d\vec{v}$ is the ion fluid velocity, and $n$ is the electron/ion density. Normalised simulation units are used: time is normalised to $\omega_{ci0}^{-1}$ space is normalised to $c/\omega_{pi0}$, charge is normalised to the proton charge $|e|$, and mass is normalised to the proton mass, where $\omega_{ci0}$ is the ion cyclotron frequency, $\omega_{pi0}$ is the ion plasma frequency, and $c$ is the speed of light in vacuum. The magnetic field is advanced in time through Faraday's Law $\frac{\partial B}{\partial t} = - \nabla \times \vec{E} $, with $\vec{E}$ calculated from Equ.~\ref{Equ1}.

In dHybrid, the ions are represented by finite sized particles moving in a 3D simulation box and are treated as kinetic particles, with their velocity updated via the Lorentz force equation. The fields and fluid quantities, such as the density n and ion fluid velocity $\vec{V_i}$, are interpolated from the particles using quadratic splines \cite[]{Decyk1996} and defined on a 3D regular grid. These quantities are then interpolated back to push the ions using quadratic splines, in a self-consistent manner. Equations are solved explicitly, based on a Boris pusher scheme to advance the particles \cite[]{Boris1970} in the hybrid approach, and on a two step LaxWendroff scheme to advance the magnetic field \cite[]{Birdsall1998, Hockney1994}. Both schemes are second-order accurate in space and time, and are space and time centred.

The present version of dHybrid uses the MPI framework as the foundation of the communication methods between processes, and the Osiris visualisation package \cite[]{fonseca2002} as the basis for all diagnostics. The three-dimensional simulation space is divided across processes; 1D, 2D and 3D domain decompositions are possible and dynamic load balancing is enabled, optimising parallel efficiency by ensuring that the computational load is similar across processors. The code can simulate an arbitrary number of particle species with arbitrary charge to mass ratios, arbitrary initial thermal velocity and drift velocity distributions, as well as arbitrary spatial configurations. Periodic boundary conditions, open boundary conditions and configurable particle injectors are used for the particles, and periodic boundary conditions are used for the fields.

Particle tracking techniques are also used in dHybrid, and are of particular relevance for the problem in hand, allowing the study of the particle acceleration mechanisms in great detail. 

Typically, a simulation is ran twice: the first time all usual diagnostics can be analysed (e.g. electric field, magnetic field, fluid phase spaces), and a special kind of diagnostics, the raw diagnostics, are produced. In these raw diagnostics, a sample of raw simulation particles are stored at given intervals, including the positions, velocities and charge. A specific set of these particles is then chosen according to specified criteria (e.g. the hundred most energetic particles, a random sample of particles). The list of particles is then supplied as input for the second run, and all the positions, velocities, electric field and magnetic field at the particle positions are stored for every iteration.

\subsection{Simulation setup}

For the problem at hand, a quasi-2D simulation setup was chosen, with one of the spatial dimensions compacted to only 5 grid cells; this setup allows for the shock structure to be resolved with higher resolution, and for the shock evolution to be followed over a longer time than would be feasible with a full 3D simulation.

The simulation frame is the shock rest frame; the CME moves faster than the surrounding solar wind, driving a shock, and thus, in the shock reference frame, the CME plasma is at rest and is represented in the simulation box by a slab of plasma in the $-x$ side of the box. 

The solar wind moves back towards the CME plasma, is present in all the simulation box, and is partially reflected at the shock front. The solar wind plasma is injected in the $+x$ side, and open boundary conditions are employed in the $x$ direction, while in the $y$ and $z$ directions periodic boundary conditions are used.

The downstream magnetic field is perpendicular to the shock normal, and simulates the solar wind magnetic field that extends as a loop from the sun surface, and is frozen in the CME plasma. The magnetic field upstream of the shock front is quasi-parallel, forming an angle of $10^\circ$ with the shock normal. This magnetic field configuration favours diffusive shock acceleration mechanisms.

The plasma kinetic to magnetic energy density ratio, $\beta$ $ = 2 nk_BT\mu_0/B^2$, (where $n$ is the plasma number density, $T$ is the plasma temperature, $\mu_0$ is the permeability of free space and $k_B$ is the Boltzman constant),  is very  sensitive to intensity of the magnetic field $|B|^2$.
However, the magnetic field intensity can be one of the hardest parameters to determine accurately \cite[]{aschwanden2004}. In-situ observational statistics \cite[]{mullan2006solar, lepping2003} show that the plasma $\beta$ in the solar wind fluctuates on either side of unity even at 1 A.U.

A super-Alfv{\'e}nic shock in the solar wind environment is modelled here, using parameters derived by \cite[]{Tsurutani2003, Wu2011, mikic2006, gary2001,aschwanden2004}. \cite[]{Tsurutani2003} described a number of different plasma parameters depending on where the CME is with respect of the ecliptic, and distance from the Sun. At distances of between 3 to 6 solar radii, and at small angles off the ecliptic, when the emerging CME has evolved from a pressure wave into a shockwave, and $\beta$ is estimated by \cite[]{Tsurutani2003} to be between 0.056 and 0.133. In our simulations we have chosen the  intermediate value of 0.08. This value is also a compromise value to aid computational efficiency which is related to the magnetic field strength. 

A CME moving in the solar wind will move into different plasma conditions as it propagates and evolves. Getting the right conditions for the process described in this paper, to create SEPs is therefore a dynamic process. In the simulations of the CME we assume that it is moving at high speed in a relatively low density solar wind making the plasma $\beta$ less than 1.

The most important parameter to maintain for these simulation is the Alfv{\'e}nic Mach numbers $M_A$, which needs to be close to $\sim$~3 for the mechanism at hand.

The choice of parameters ensures a behaviour that is identical to the real shock scenario, while guaranteeing that the simulation is feasible and numerically stable. For the CME plasma (where $\beta = 0.05$) the density is $n_{CME} = 10^4\, cm^{-3}$, and the ion temperature is $T_i = 0.1 \,eV$, while for the solar wind (where $\beta = 0.08$) the density is $n_{sw} = 1000 cm^{-3}$, and $T_i = 2 \,eV$. The solar wind is drifting towards the CME at $190\, kms^{-1}$, equivalent to $M_A = 2.75$, for a background magnetic field of $100\, nT$.

Results are presented in simulation units, with the density normalized to $n_0 = 10 cm^{-3}$, the distance to $c/\omega_{pi0} = 71.96\, km$, the time to $\omega^{-1}_{ci0} = 3.69\, s$, the velocity to $v_{A0} = 19.5 kms^{-1}$, the magnetic field to $B_0 = 2.825\, nT$, and the electric field to $B_0 v_{A0} = 0.0551\,mV/m$.

The simulation box size is $32 \times 16 \times 0.3125\,(c/\omega_{pi0})^3$, equivalent to $116.56 \times 58.13 \times 1.14\, (r_{Lsw})^3$ (solar wind Larmor radius), with $1024 \times 512 \times 5$ grid cells, corresponding to a grid cell size of $0.03125\,c/\omega_{pi0} = 0.11\,r_{Lsw}$. The simulation is run up to 312000 iterations, with a time step of $1.28 \times 10^{-5}\,\omega^{-1}_{ci0}$ , equivalent to $7.28 \times 10^{-5}\,T_{csw}$ (ion cyclotron periods of the solar wind), yielding a total simulation time of $4.08\,\omega_{ci0} = 23\,T_{csw} = 15\,s$.

\section{RESULTS}
\subsection{Wave generation upstream of the shock}

\begin{figure*}
        \includegraphics[width=1\textwidth]{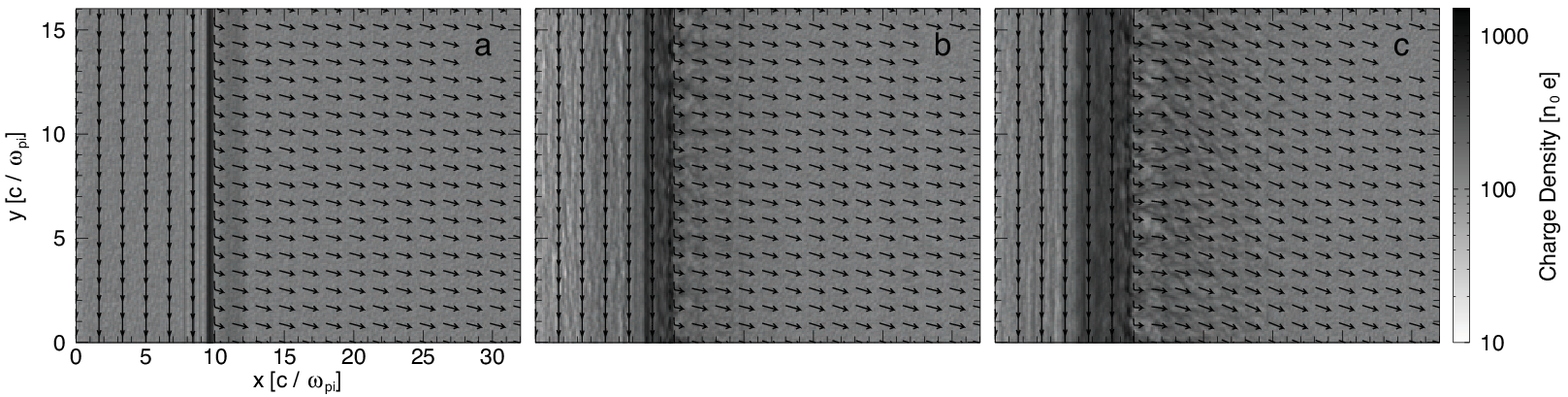}
\caption{Solar wind charge density and magnetic field vectors for times a) $t= 0.4\,\omega^{-1}_{ci0}$ , b) $t = 1.36 \,\omega^{-1}_{ci0}$ , and c) $t = 2.32 \,\omega^{-1}_{ci0}$.}\label{FIG1}
\end{figure*}

Figure~\ref{FIG1} shows the charge density of the solar wind super-imposed with the magnetic field lines for three distinct moments in time. The shock front is defined at $x = 10$ by the density jump between the solar wind upstream and the CME plasma downstream, as well as by the jump in direction of the magnetic field; the solar wind plasma reflected at the shock is strongly modulated, in the upstream region, and the magnetic field intensity does not suffer dramatic changes ($\delta B/B <<1$), although the field direction varies slightly. The plasma density perturbations upstream of the shock become more turbulent with time, which is an indication of wave breaking that produces turbulence in the non-linear regime.

By looking at the shock behaviour, it is patent that the solar wind plasma reflected at the shock front and propagating upstream drives electromagnetic waves in the upstream region. Figure~\ref{FIG2} shows the time evolution of the $v_x x$ phase space for the solar wind plasma, with the super-imposed electric field intensity line-out along the shock direction, and shows that a wave is formed by the interaction of the two counter streaming ion populations. The same oscillations are observed in the magnetic field (figure~\ref{FIG3}) indicating the presence of an electromagnetic wave. Also from figure~\ref{FIG3} , it is seen that $\delta E/E >> \delta B/B$, and that oscillations occur in the $y$ and $z$ components of both the Electric field and the Magnetic field, while there is a smaller amplitude oscillation of the $x$ component of the Electric field.

Measuring the wavelength of the wave yields $\lambda = 3\,c/\omega_{pi0}$, so that $k = 2.09\,\omega_{pi0}/c$, and measuring the propagation velocity of the wave front yields $v = 6.2 v_{A0}$, which is consistent with an Alfv{\'e}n wave with frequency $\omega = 12.99\,\omega_{ci0}$, in the simulation reference frame. This wave is actually supported by the reflected solar wind plasma, and in the reference frame moving with this plasma the wave actually propagates in the $-x$ direction with the Alfv{\'e}n velocity of $3.5397\, v_{A0}$, yielding a frequency $\omega = 7.41\,\omega_{ci0}$. The wave is then a rotating elliptically polarized Alfv{\'e}n wave propagating along $x$ with main components in the $y$ and $z$ directions, and with a smaller ($\delta E_x << \delta E_y $ and $\delta E_x << \delta E_z$) electrostatic component directed along the propagation axis $x$.

Wave formation due to counter-streaming super-Alfv{\'e}nic ion populations is a known effect \cite[]{McKenzie1982, Gordon1999, Zank2000, Ng2003, Rice2003} different modes can be excited, from MHD modes \cite[]{Lee1983}, to kinetically driven Alfv{\'e}n waves, and to purely growing instabilities, relevant for Cosmic Ray acceleration mechanisms \cite[]{Lucek2000, Bell2004}. Our results show an elliptically polarised Alfv{\'e}n wave, and include also an electrostatic component along the $x$ direction. This component is due to the quasi-parallel magnetic field configuration that increases the complexity of the configuration, in comparison with the parallel magnetic fields usually assumed in the theoretical models. The wavelengths and growth rates are compatible with the instabilities described by \cite[]{Lucek2000, Bell2004}. For this instability, small wavelengths grow with time until a maximum wavelength is reached, beyond which the instability saturates. The quasi-linear MHD theory of the instability predicts a growth rate of $\gamma_{max} = \zeta\, v^2_{sh}/(2\,v_A\,r_{Lsp1})$ for the fastest growing wave number $k_{max} = \gamma_{max} v_A^{-1}$ , with $\zeta = B_0\,j\,r_{Lsp1}\,\rho^{-1}\,v_{sh}^{-2}$, in the non-relativistic regime. The unstable wave vector range is $ 1< kr_{Lsp1} < \zeta v_{sh}^2v_A^{-2}$. The instability works for parallel and quasi-parallel shocks, as inour case, and when one of the species is unmagnetized and the other species is magnetized. Here, $v_{sh}$ is the relative velocity between the two plasma species, $\rho$ is the mass density for the background (magnetized) species, $j$ is the current density of the unmagnetized species,$r_{Lsp1}$ is the Larmor radius for the unmagnetized species, and $v_A$ is the Alfv{\'e}n velocity.

In our case, the ions reflected at the shock front get unmagnetized due to scattering, while the ions that are streaming towards the shock front are magnetized (they are streaming along a quasi-parallel magnetic field with a relatively low temperature). 

The ion Larmor radius of $r_{Lsp1} \sim 0.73\, c/\omega_{pi0}$ can be measured directly in the simulation, but the density ratio $n_{sp1}/n_{sp0}$ of the two counter-streaming ion populations, controlling the parameter $\zeta$ through the current $j$ and mass density $\rho$, varies strongly during the simulation. 

This is not accounted for in the theoretical model, that assumes a constant current driving the instability, an isotropic, or power law, particle distribution for the unmagnetized species, and propagation parallel to the magnetic field \cite[]{Bell2004}. Since the current driving the instability is not constant, the propagation is quasi-parallel, and the particle distribution, at the spatial lengths considered, is not isotropic (c.f. figures~\ref{FIG2} and \ref{FIG4}), only an order of magnitude estimation can be done for the theoretical values of $\gamma_{max}$ and $k_{max}$. For $n_{sp1}/n_{sp0} \sim 0.08$, as in the early stages of the simulation, a growth time of $\gamma_{max} \sim 7\,\omega_{ci0}^{-1}$ and a wave number of $k_{max} \sim 2\, \omega_{pi0}/c$ can be estimated.

\section{Particle acceleration}

\begin{figure*}
        \includegraphics[width=1.0\textwidth]{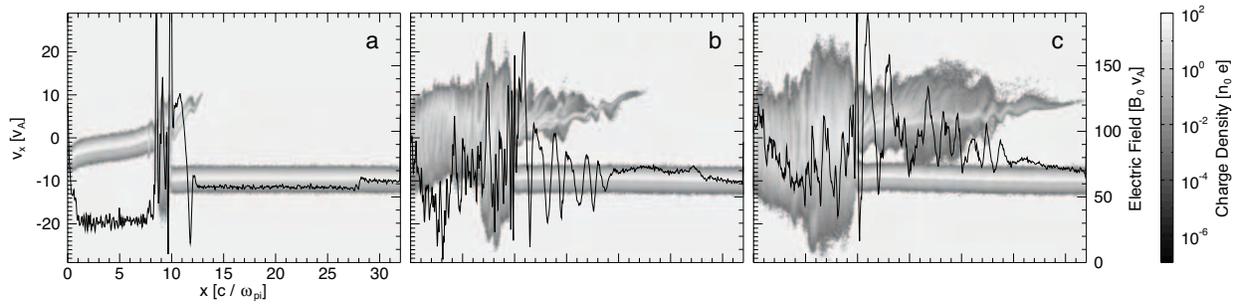}
\caption{Solar wind $v_x x$ phase space and electric field lineout along the $x$ direction in the center of the simulation box ($y = 8\, c/\omega_{pi0}$) for times a) $t = 0.4\,\omega^{-1}_{ci0}$, b) $t = 1.36 \,\omega^{-1}_{ci0}$, and c) $t = 2.32 \,\omega^{-1}_{ci0}$.}\label{FIG2}
\end{figure*}

\begin{figure*}
        \includegraphics[width=0.5\textwidth]{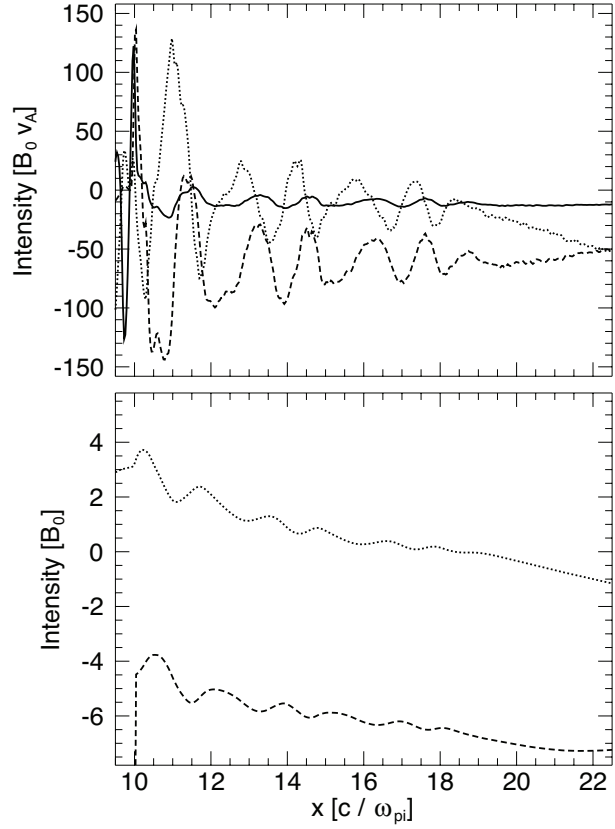}
         \centering
\caption{Electric field lineout (top panel), and magnetic field lineout (bottom panel), at time $t = 1.36 \,\omega^{-1}_{ci0}$ . The lineout is along the $x$ direction at $y = 8 c/\omega_{pi0}$; the full line represents the $x$ component (electric field), the dashed line represents the $y$ component (electric field and magnetic field), and the dotted line represents the $z$ component (electric field and magnetic field.}\label{FIG3}
\end{figure*}

\begin{figure*}
        \includegraphics[width=1.0\textwidth]{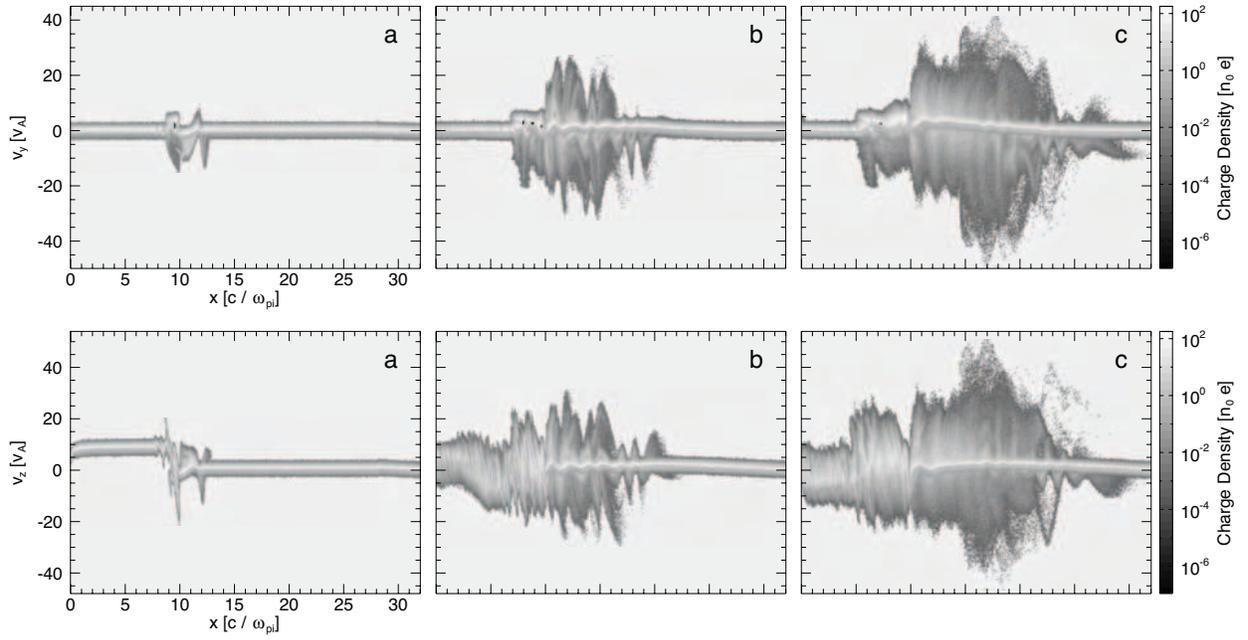}
\caption{Solar wind vy x phase space (top panel), and solar wind $v_z x$ phase space (bottom panel), for times a) $t = 0.4\,\omega^{-1}_{ci0}$ , b) $t = 1.36 \,\omega^{-1}_{ci0}$ , and c) $t = 2.32 \,\omega^{-1}_{ci0}$.}\label{FIG4}
\end{figure*}

A significant number of particles reflected at the shock front are seen to interact with the previously formed waves and accelerate. From the inspection of figure~\ref{FIG2} and figure~\ref{FIG4}, showing the $v_x x$, $v_y y$, and $v_z z$ phase spaces, it is seen that the energy gain is mostly in the $y$ and $z$ directions, that is, in the directions perpendicular to the shock propagation.

Another interesting observation is that a part of the particle population that is streaming in the $+x$ direction is being reflected back to the shock at $x \sim 15\, c/\omega_{pi0}$ for later times, visible in frame c) of figure~\ref{FIG2} , where some of the particles in the upper branch of the phase-space plot have negative velocities at that point. Also, the particles with the greatest perpendicular velocities up to $t = 2.53\, \omega_{ci0}^{-1}$ (frame c), are moving away from the shock front.

At later times, $t \sim 4 \, \omega_{ci0}^{-1}$ , in the simulation,there is indication of a new group of particles gaining energy in the region of the shock front around $x \sim 10\, c/\omega_{pi0}$.

\begin{figure}
        \includegraphics[width=0.5\textwidth]{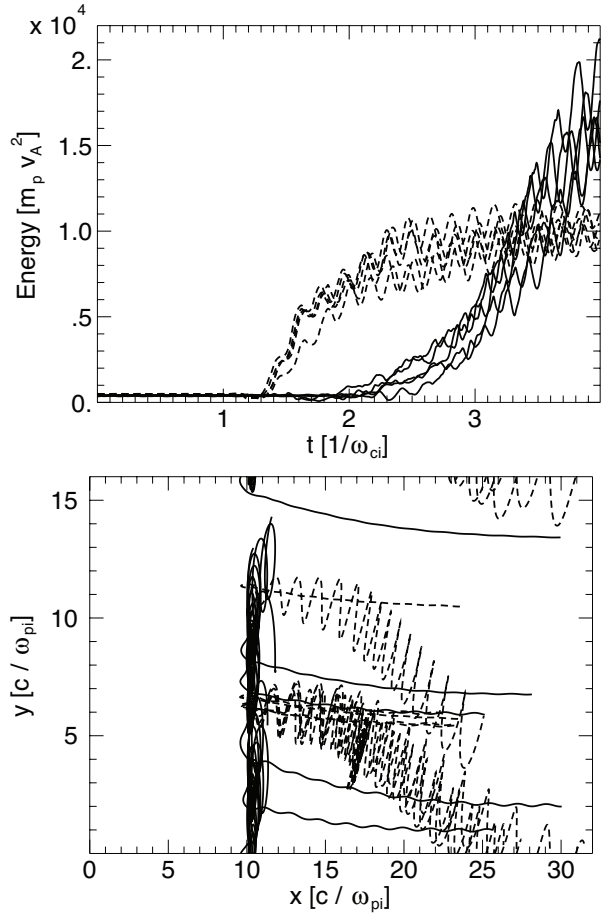}
         \centering
\caption{Evolution of the particle energy (top panel), and particle trajectory (bottom panel). Dashed lines represent the particles accelerated earlier in time that drift away from the shock front (group 1), full lines represent particles drifting along the shock front (group 2).}\label{FIG5}
\end{figure}

Looking directly at the kinetics of the most energetic particles provides insights on the physical processes that dominate particle acceleration. The particle tracks for the top-80 most energetic particles in the simulation reveals two distinct groups of particles. Figure \ref{FIG5} shows the five most energetic particles from each of these two groups. Particles from group 1 are accelerated very early in time, and the acceleration occurs in two distinct phases: a strong energy gain around $1.3 < t < 2.0~\omega_{ci0}^{-1}$, and a weaker increase in energy from $t \sim 2 ~\omega_{ci0}^{-1}$ onwards. These are the particles that move away from the shock front, reaching energies up to $22.5$ times their initial energy (dashed lines, figure~\ref{FIG5}), that are in a zone dominated by the Alfv{\'e}n wave, and causing the energetic particle population seen in figure~\ref{FIG4} frames b) and c).

Particles from group 2 (figure~\ref{FIG5}) start gaining energy only later, around $t \sim 2\, \omega_{ci0}^{-1}$. This kind of behaviour for Interplanetary Shocks, with two distinct phases, is predicted by \cite[]{Lee1983}, who presents a model for turbulence enhancement due to counter-streaming ion populations, generation of waves in the upstream media, and DSA acceleration of particles due to this turbulence. 

At $t \sim 2\omega_{ci0}^{-1}$, the fields still preserve a wave-like structure in regions away from the shock, as can be seen in the electric field lineout, $17 < x < 25~c/\omega_{pi0}$, in figure~\ref{FIG2} frame c). Spatial regions near the shock front, however, start to exhibit turbulence, mostly visible in the density plot of figure~\ref{FIG1} frame c), upstream of the shock front. This turbulence intensifies with time and is the reason why particles in group 2 are trapped in the shock front and start to shock drift, gaining energies up to $50$ times their initial energy in
$t \sim 2\omega_{ci0}^{-1} \sim 7.5$s. The total energy gain for these particles is $\sim 110$ times the maximum energy that a particle could gain be simply crossing the shock front once. 

With an energy increase in time that is approximately quadratic (figure~\ref{FIG5}) it would be possible for a typical $1~keV$ proton to reach an energy of $\sim 200~MeV$ in around $10$ minutes time, if the energy gain could be sustained for that period of time.

\begin{figure*}
        \includegraphics[width=1.0\textwidth]{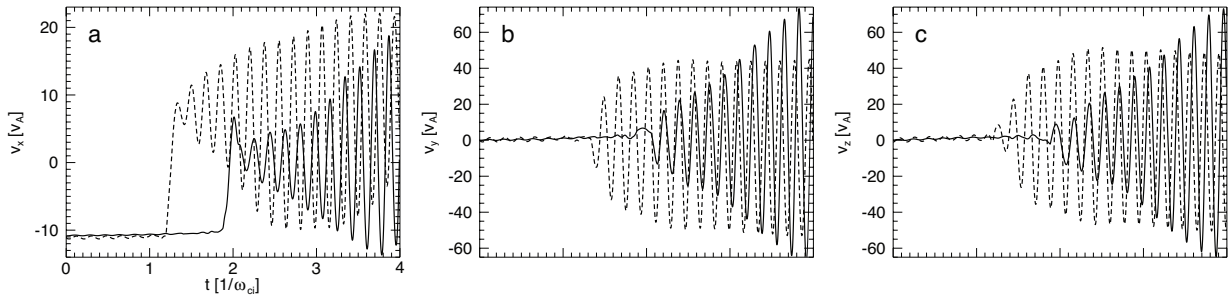}
\caption{Velocity evolution in time for the most accelerated particle from group 1, and group 2, identified in figure 5. Frame a) represents $v_x$, frame b) represents $v_y$, and frame c) represents $v_z$.}\label{FIG6}
\end{figure*}

Figure~\ref{FIG6} shows the velocity components of the most accelerated particle in group 1 (dashed), and the most accelerated particle in group 2 (full line). The fundamental difference between the two particles is that particle 1 (from group 1) has a positive $v_x$ velocity, and traverses the most efficient acceleration zone, situated in front of the shock, very quickly, gaining most of its energy in a time interval $t\sim 0.7$. The velocity increase in this period is, however, approximately linear ($1.3 < t <2$, figure~\ref{FIG6}). Particle 2 (from group 2) exhibits the same behaviour: the mean velocity increases linearly in the time interval 
$2 < t < 4$, although the acceleration is more efficient, due to the initial $v_x \sim 0$ velocity. The result is that particle 2 drifts along the shock front, in the upstream region, while particle 1 follows the wave propagating away from the shock.

The velocity profile increase seen in figure~\ref{FIG6}, in the acceleration phases, is consistent with the picture of a particle trapped in a circularly polarised Alfv{\'e}n wave. In the simplest form, we consider a zero-order magnetic field parallel to the shock normal, $B_x$, and an Alfv{\'e}n wave with amplitude $A_0$ and components in $\vec{e_y}$ and $\vec{e_z}$, with an electric field
component $\vec{E} = A_0 \,[\, \cos\,(kx - \omega t)\,\vec{e_y} - \sin\,(k x -  \omega t)\,\vec{e_z}\,] $, and a magnetic field component $ \vec{B} = A_0\, [\, \sin \,(k x - \omega t)\,\vec{e_y} + \cos\,(k x - \omega t)\,\vec{e_z}\,] $. An ion can be trapped in this wave if, in zeroth-order, $v_x = \omega/k + \omega_{cx}$
with $\omega_{cx} = q\,B/m$; solving the single particle motion using the Lorentz force equation and using the above trapping condition yields $v_y(t) = K_1 \cos (\omega_{cx}\,t)\,t$ and$v_z(t) = -K_1 \sin\, (\omega_{cx}\,t)\,t$
with $K_1$ constant.

The above picture recovers the behaviour seen in figure 6 for the perpendicular velocity components $v_y$ and $v_z$, and does not explain any energy gain along the magnetic field in the $x$ direction. If we refer again to figure~\ref{FIG3}, we can see that the simple assumptions made are over-simplistic, and instead an electrostatic wave component would have to be considered, along with different electric field and magnetic field amplitudes, and also finite values for the static magnetic field components. The model describes the main qualitative features of the acceleration well, while the quantitative behaviour in the much more realistic simulation scenario is more complex.

\section{CONCLUSIONS}

We have presented 2D hybrid simulation results of a quasi-parallel shock, with realistic shock parameters. In the shock reference frame used in the simulation, the solar wind plasma flows along the quasi-parallel magnetic field, hits the Coronal Mass Ejection plasma, and is scattered. The upstream population of scattered ions induces the formation of an electromagnetic Alfv{\'e}n wave. In a completely self-consistent picture, the Alfv{\'e}n waves create turbulence upstream and accelerate a significant population of ions.

The results presented are qualitatively different from those provided by the usual MHD simulation techniques: the shock propagation is followed on a different time-scale, relevant for the ion dynamics, and shorter than the typical MHD simulation time-scale.

Also the simulation is localised in space in comparison with MHD simulations than can model the global behaviour of a CME. The detailed spatial and temporal resolution attained results in a much more complete physical picture, in which there are electromagnetic waves propagating due to the counter-streaming ion populations in the upstream region, and in which particles are accelerated in two distinct phases. 

The shock propagates for a time interval $T = 4\,\omega_{ci0}^{-1} \sim 15~s$, and the most accelerated particles start gaining energy at $t = 2\, \omega_{ci0}^{-1}$ from an initial thermal distribution. The energy gain is approximately quadratic in time up to $50$ times the initial energy, meaning that if a part of these ions could be trapped for periods of time of $\sim 10$ minutes, the energy gain could lead to particles with $\sim 200\,MeV$, consistent with observations.

The crucial question of whether the energy gain is sustainable for long periods of time requires further investigation. Shock-drift theory dictates that, for a particle in a perpendicular shock, the energy gain is proportional to the magnetic field compression ratio which, depending on the shock strength, means an energy gain of up to 5 times the initial energy of a particle. In this case, due to the wave structure present at the shock front, the observed energy gain is much greater, going up to 50 times the initial energy of a particle that is shock drifting. Particles will always drift away from the shock front and that means that for further acceleration a diffusive shock mechanism, driving the particles back to the shock, is necessary.

Our simulations provide evidence of particles being reflected back to the shock (cf. figure~\ref{FIG5}), suggesting that diffusive shock acceleration (DSA) actually occurs in the configuration considered, and indicating that the hybrid simulation model is capable of correctly modelling the mechanism. While a better characterisation of the DSA mechanism can be done because particle kinetics can be directly observed, a complete understanding of the mechanism involves modelling the shock propagation for times longer than those presented in this paper.

\section{Acknowledgements}
This work was supported by the European Research Council (ERC-2010-AdG Grant 267841) and by Funda\c{c}\~{a}o para a Ci{\^e}ncia e a Tecnologia (FCT/Portugal) under grant SFRH/BD/17750/2004. The simulations presented in this paper were produced using the IST Cluster (IST/Portugal). The authors would also like to thank the U.K. STFC Center for Fundamental Physics for it's support.

{\it Facilities:} \facility{dHybrid}


\bibliography{SEPCMEbib3}{}
\bibliographystyle{apalike}

\clearpage



\end{document}